\documentclass[conference]{IEEEtran}
\IEEEoverridecommandlockouts
\usepackage[]{collab}
\usepackage{enumitem}
\usepackage{balance}
\usepackage{xcolor}
\usepackage{setspace}
\usepackage{subfigure}
\usepackage{multirow}
\usepackage{multicol}
\usepackage{tabularx}
\usepackage{siunitx}
\usepackage{float}
\usepackage{color}
\usepackage{hyperref}
\usepackage{diagbox}
\usepackage{caption}
\usepackage{makecell}
\usepackage{colortbl}
\usepackage{tcolorbox}
\usepackage{mdframed}
\usepackage{listings}
\usepackage{booktabs}
\usepackage[utf8]{inputenc}
\usepackage{calligra}
\usepackage[normalem]{ulem}
\usepackage{indentfirst}
\usepackage{url}
\usepackage{soul}
\usepackage{nicematrix}
\usepackage{fontawesome5}
\usepackage{textcomp}
\usepackage{stfloats}
\usepackage{verbatim}
\usepackage{graphicx}
\usepackage{amsmath,amsfonts}
\usepackage[linesnumbered,ruled,vlined,noend]{algorithm2e}
\usepackage{array}
\usepackage[utf8]{inputenc}

\collabAuthor{jj}{blue}{Junjie}
\collabAuthor{gb}{magenta}{Guangba}
\collabAuthor{zh}{purple}{Zhihan Jiang}
\collabAuthor{jz}{magenta}{Jiazhen Gu}
\collabAuthor{zb}{teal}{Zhuangbin}
\collabAuthor{jc}{googlered}{JC}
\collabAuthor{yj}{orange}{Yujie Huang}
\collabAuthor{yl}{orange}{Yulun Wu}

\newcommand\nm{LLMPrism\xspace}
\newcommand\platform{Platform-X\xspace}

\newcommand{\ie}{{\em i.e.},\xspace}
\newcommand{\eg}{{\em e.g.},\xspace}

\newcommand{\boxmargin}{1mm}

\newtcolorbox{myboxa}[2][]{
    colback=gray!10!white,
    colframe=black, enhanced,
    attach boxed title to top left={yshift=-2mm,xshift=5mm},
    title=#2,#1
}
\newtcolorbox{myboxb}[2][]{
    boxsep=3pt,
    left = \boxmargin, right = \boxmargin, top = \boxmargin, bottom = \boxmargin,
    title={#2},#1
}
\newtcolorbox{myboxc}{
    colback=gray!15!white,
    arc = 0pt, outer arc = 0pt,
    boxsep=0pt, left = 3pt, right = 0pt, top = 0pt, bottom = 0pt, 
    leftrule=3pt, bottomrule=0pt,toprule=0pt, rightrule=0pt,
    left = \boxmargin, right = \boxmargin, top = \boxmargin, bottom = \boxmargin
}
\newtcolorbox{myboxd}{
    colback=gray!10,
    colframe=black,
    width=\columnwidth,
    arc=1mm, auto outer arc,
    boxrule=0.5pt,
}

\definecolor{myyellow}{HTML}{FFF2CC}
\newcounter{finding}

\definecolor{myyellow}{HTML}{FFF2CC}
\newcounter{insight}

\newcounter{challenge}

\definecolor{mygreen}{HTML}{AFCFA5}
\newcounter{opportunity}

\begin{document}

\thispagestyle{plain}
\pagestyle{plain}

\title{\nm: Black-box Performance Diagnosis for Production LLM Training Platforms}

\author{\large Zhihan Jiang\IEEEauthorrefmark{2}, Rui Ren\IEEEauthorrefmark{3}, Guangba Yu\IEEEauthorrefmark{2},Yulun Wu\IEEEauthorrefmark{2}, Wenwei Gu\IEEEauthorrefmark{2}, Yichen Li\IEEEauthorrefmark{2},  Yujie Huang\IEEEauthorrefmark{2} \\ Cong Feng\IEEEauthorrefmark{3}, Zengyin Yang\IEEEauthorrefmark{3}, Yongqiang Yang\IEEEauthorrefmark{3}, Michael R. Lyu\IEEEauthorrefmark{2}\\
\IEEEauthorblockA{
\IEEEauthorrefmark{2}The Chinese University of Hong Kong, Hong Kong SAR, China \\
\IEEEauthorrefmark{3}Computing and Networking Innovation Lab, Huawei Cloud Computing Technology Co., Ltd, China\\
}
}

\maketitle

\begin{abstract}

Large Language Models (LLMs) have brought about revolutionary changes in diverse fields, rendering LLM training of utmost importance for modern enterprises.
To meet this demand, multi-tenant large-scale LLM training platforms have been built to offer LLM training services.
Nevertheless, due to the complexity and synchronous nature of LLM training process, performance issues occur frequently and can result in substantial resource wastage.
The limited visibility from the perspective of platform providers impedes existing profiling methods and poses challenges to the monitoring and diagnosis of the performance of LLM training jobs.
For the first time, this paper proposes the utilization of underlying network flow data to reconstruct the training timelines of jobs based on the distinct characteristics in the LLM training procedure.
We design \nm, the first black-box performance diagnosis system for LLM training platforms.
By progressively recognizing LLM training jobs, identifying their parallelism strategies, and reconstructing the training timelines, \nm achieves non-intrusive, lightweight, and continuous monitoring of LLM training systems.
Leveraging this monitoring capability, it further effectively diagnoses potential performance issues.
Since Oct. 2024, \nm has been deployed on our large-scale production \platform, in which the evaluations and deployment experiences demonstrate that \nm can achieve accurate timeline reconstruction with an error within 0.3\% and effectively diagnose various performance issues.

\end{abstract}


\section{introduction}

Large language models (LLMs) have emerged as transformative technologies across various domains~\cite{jin2024comprehensive,liu2023scalable,li2024exploring,jiang2024lilac,li2024go,li2025coca}.
Their remarkable performance is predominantly attributed to the scaling law~\cite{kaplan2020scaling}, which indicates a strong correlation between model capacity and both model size and training data volume. For example, contemporary models such as DeepSeek-R1~\cite{guo2025deepseek} comprise up to 671 billion parameters.

Achieving state-of-the-art capabilities in LLMs requires extensive efforts in training or tuning, demanding substantial computational resources~\cite{he2023unicron,DBLP:conf/sc/AnBCCDDDDGGGGFH24}.
Illustratively, the Llama3-405B model was trained using 16,384 H100 GPUs for 54 days~\cite{dubey2024llama}.
To support such resource-intensive workflows, leading IT enterprises have developed multi-tenant platforms, such as Amazon SageMaker~\cite{amazon-ai} and Google Vertex AI~\cite{googlevertex}, adopting a machine-as-a-service (MaaS) paradigm. These platforms enable tenants to rent GPU clusters for distributed training while concealing sensitive configuration details (\eg machine counts and parallelism strategies) for privacy reasons.
This opacity results in a black-box view for providers, complicating performance diagnostic and optimization~\cite{dongevolution,liu2023prism,huang2024faultprofit}.

The scale and complexity of these training jobs, coupled with synchronization requirements at training-step boundaries, often lead to performance degradation~\cite{wu2024falcon,jiang2024megascale,kokolis2024revisiting,jiang2025l4}.
Recent research has identified that these slowdowns stem from various factors such as network congestion~\cite{jin2024communication}, resource contention~\cite{feng2024optimus}, and thermal throttling~\cite{patel2024accelerating}, causing resource wastage and extended training durations~\cite{wu2024falcon,cui2025xputimer}.
Although various profiling tools~\cite{cui2025xputimer,HTA,strobelight} have been proposed to track training processes, their effectiveness diminishes in black-box multi-tenant platforms.
These tools typically require intrusive code or configuration modifications, or face compatibility issues with different frameworks, restricting their applicability (details shown in §~\ref{sec:motivation}).

Consequently, providers of multi-tenant training platforms urgently require a lightweight, long-running diagnostic framework operable in black-box scenarios.
Among various performance issues, network-related problems constitute a significant portion~\cite{wu2024falcon}, primarily because distributed LLM training inherently requires substantial inter-machine communication.
Following practices common in traditional data centers, most LLM training platform providers have implemented network monitoring solutions, such as ERSPAN~\cite{ERSPAN}, which leverages packet mirroring at the switch level to capture comprehensive network flow information.
These network monitors independently record detailed communication histories throughout the training process, imposing minimal impacts on running job performance.
Despite the availability of this extensive network flow data, effectively analyzing it to diagnose upper-layer performance issues of LLM training jobs remains an unexplored and challenging approach.

To bridge this gap, we propose leveraging underlying network data to reconstruct detailed LLM training timelines.
This approach transforms the original black-box view into white-box, enabling precise diagnosis of performance issues within large-scale training platforms.
Specifically, this paper introduces \nm, the first network flow-based framework designed for performance diagnosis in LLM training platforms, offering non-intrusive monitoring and diagnostic capabilities while incurring minimal performance impact.

The design of \nm is based on three key characteristics inherent to LLM training communications: \emph{spatial patterns}, \emph{temporal patterns}, and \emph{distinctive parallelism features}, which are elaborated in §~\ref{sec:motivation}.
Building upon these insights, \nm utilizes network flow data to progressively reconstruct training timelines and diagnose potential performance issues.
Initially, \nm recognizes individual LLM training jobs within the platforms hosting hundreds of jobs based on spatial communication patterns.
Subsequently, it robustly identifies the parallelism strategies employed by each job through their distinctive communication characteristics.
Then, for each GPU, \nm reconstructs the precise training timeline by analyzing temporal communication patterns.
Finally, through multidimensional degradation analysis of these reconstructed timelines, \nm effectively pinpoints and diagnoses performance issues within LLM training jobs.

\nm has been deployed in our production multi-tenant LLM training \platform, supporting LLM training services for hundreds of internal users and partner organizations since Oct. 2024.
Evaluations on real-world large-scale LLM training jobs and our deployed experiences, demonstrate that \nm reliably achieves 100\% accuracy in identifying training jobs and their parallelism strategies, even using network flows over just a one-minute interval.
Furthermore, \nm maintains a reconstruction error rate below 0.3\% for training timelines, aiding precise performance diagnosis and network issue identification.

In summary, the main contributions of this paper are below:
\begin{itemize}[leftmargin=*, topsep=0pt, parsep=0pt]
\item To the best of our knowledge, this work is the first to highlight the feasibility of reconstructing LLM training timelines from underlying network flow data.
\item We implement and deploy \nm, the first non-intrusive network flow-based diagnostic framework for LLM training.
\item We conducted evaluations and share our deployment experiences, validating \nm's effectiveness and practicality.
\end{itemize}

\section{Background}

\subsection{Parallelism Strategies in LLM Training}
\label{background:training}

Large-scale distributed training of LLMs typically employs thousands of heterogeneous accelerators (\eg GPUs or NPUs; hereafter uniformly referred to as GPUs for simplicity) interconnected via high-speed networks such as Remote Direct Memory Access (RDMA) over Converged Ethernet (RoCE)\cite{kaur2013rdma}.
To facilitate parallel training, several strategies are used, as demonstrated in Fig.~\ref{fig:background}: \emph{Data Parallelism (DP)} creates model replicas across GPUs, partitioning the dataset into mini-batches, thus requiring synchronization communication~\cite{shoeybi2019megatron}; \emph{Pipeline Parallelism (PP)} distributes model layers across multiple GPUs, processing micro-batches within each step~\cite{rajasekaran2024cassini};
and \emph{Tensor Parallelism (TP)} partitions operations across GPUs, incurring substantial communication volume~\cite{zheng2022alpa}, and is thus typically limited to intra-node.
We primarily focus on DP and PP, which require cross-node communications, making them susceptible to network issues.

Once the LLM training job begins, the parallelism strategy remains fixed throughout the training process.
Subsequently, the training proceeds step-by-step.
In each step, GPUs within the same PP group perform calculations in a pipeline manner, where each GPU processes a stage of the model. For each mirco-batch, during forward propagation, each GPU sends intermediate activations to the subsequent GPU, while during backward propagation, gradients are transmitted back to the preceding GPU.
At the end of each training step, GPUs within the same DP groups engage in collective DP communications to synchronize gradients and update model parameters.

\subsection{LLM Training Network Monitoring}
\label{sec:background_network_monitoring}

\begin{figure}[t]
    \centering
    \includegraphics[width=\linewidth]{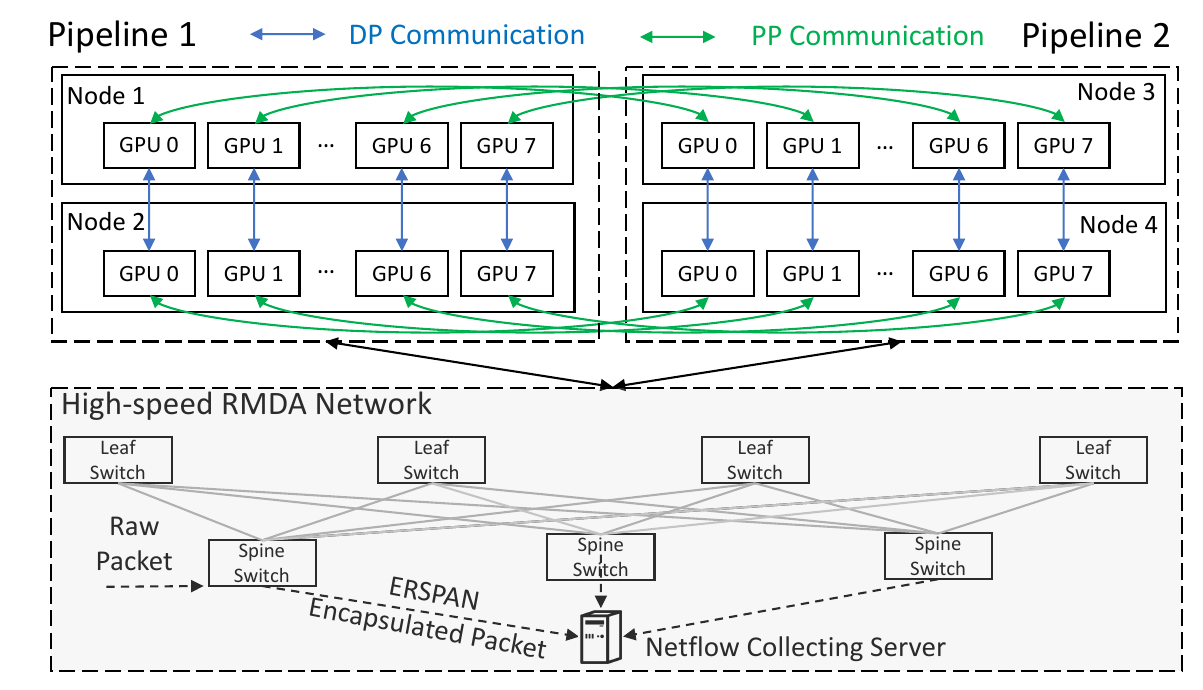}
    \caption{The training parallelism and network monitoring.}
    \label{fig:background}
\end{figure}

Similar to conventional data centers, providers of LLM training platforms typically deploy specialized tools (\eg Meta ROCET~\cite{gangidi2024rdma}) at the switch level, 
A prominent example is the Encapsulated Remote Switch Port Analyzer (ERSPAN)~\cite{ERSPAN}, which enables remote monitoring by mirroring network traffic from a source port to an individual storage server, as shown in the lower part of Fig.~\ref{fig:background}. 
This non-intrusive and independent approach minimizes interference with the network and associated workloads, as it requires only mirroring the relevant portions of network packets, enhancing its practicality and versatility.
Specifically, the collected network flows typically include \emph{flow start time}, \emph{source address}, \emph{destination address}, \emph{involved switches}, \emph{flow size}, and \emph{flow durations}.
Despite the common availability of these underlying network measurements, they remain underutilized in current LLM training platforms.

\section{Motivation}
\label{sec:motivation}

Large-scale LLM training is highly susceptible to slowdown due to stringent synchronization requirements, which can lead to substantial wastage of computational resources and training time~\cite{wu2024falcon,jiang2024megascale,zhang2023opt}.
Consequently, effective performance monitoring and analysis become crucial for LLM training jobs.
To achieve this, various profiling tools~\cite{HTA,strobelight,cui2025xputimer} have been proposed, which continuously track the operational durations throughout the training process.
Nevertheless, existing profiling approaches present following limitations, particularly within the context of LLM training service platforms:
\begin{itemize}[leftmargin=*, topsep=0pt, parsep=0pt]
\item \emph{Intrusiveness and Compatibility.} Current profilers require intrusive modifications to user/framework code or configurations and often have compatibility constraints, rendering them impractical for adoption in LLM training platforms.
\item \emph{Performance Overhead.} Profiling tools inevitably introduce additional overhead, negatively affecting the original training efficiency. For instance, XPUTimer~\cite{cui2025xputimer} incurs a 5\% performance degradation in jobs utilizing 3D parallelism.
\item \emph{Visibility Constraints.} Due to tenant privacy concerns, users typically refrain from sharing job configurations and profiling data with platform providers, leading to severely limited visibility for performance analysis.
\end{itemize}

Collectively, these challenges significantly impede LLM training platform providers' abilities to diagnose job slowdown under such black-box view.
Among the various fail-slow issues, cross-node network faults are notably predominant.
As highlighted in \cite{wu2024falcon}, network congestion accounts for more than 81\% of slowdown cases in large-scale training jobs.
Hence, platform providers commonly deploy monitors at the network switch layer, such as ERSPAN~\cite{ERSPAN} and ROCET~\cite{gangidi2024rdma}, to identify potential network issues.
Nevertheless, existing diagnostic methodologies fail to fully leverage available network flow data, resulting in insufficient performance analysis.

To overcome these limitations, for the first time, we propose the utilization of network flow data to reconstruct LLM training timelines.
This novel approach shifts the analysis from a black-box to a transparent, white-box perspective, enabling continuous monitoring and comprehensive performance diagnosis for LLM training jobs.
Importantly, this method is non-intrusive and introduces minimal performance impact on original training jobs. 
This is feasible based on three observed distinct characteristics inherent in LLM training procedure:
\begin{itemize}[leftmargin=*, topsep=0pt, parsep=0pt]
\item \emph{Spatial Communication Patterns.} Communications during LLM training exhibit spatial stability, with interactions strictly confined to GPUs within the same DP or PP groups.
\item \emph{Temporal Communication Patterns.} Communications traffics demonstrate clear temporal periodicity, as the training steps keep cycling and the workload remains stable.
\item \emph{Distinctive Parallelism Communication Features.} Different communication modes (DP and PP) display uniquely identifiable characteristics, enabling differentiation between them.
\end{itemize}

\begin{figure*}[t]
    \centering
    \includegraphics[width=0.96\linewidth]{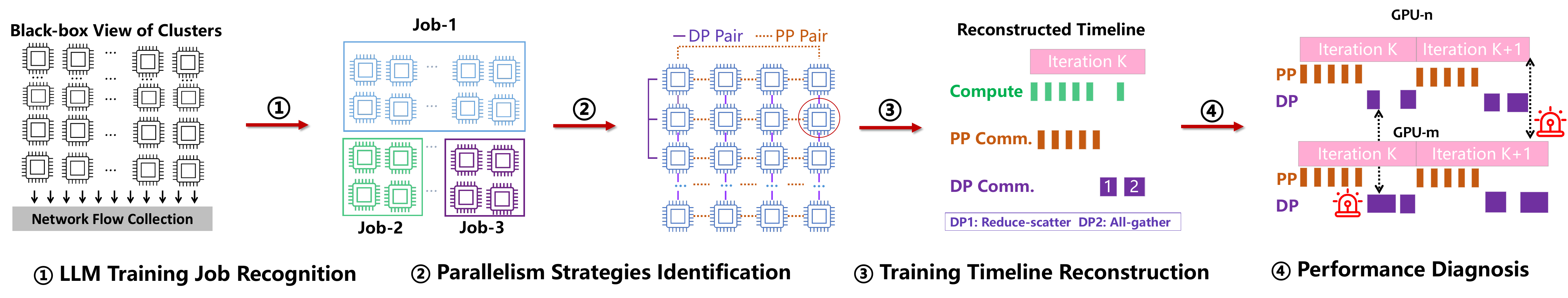}
    \caption{The overall framework of \nm.}
    \label{fig:framework}
\end{figure*}

\section{Methodology}

To non-intrusively monitor and diagnose the performance of LLM training jobs, we propose \nm, the first network flow-based diagnosis framework for large-scale LLM training platforms.
The collected network flow data follows the format introduced in Sec.~\ref{sec:background_network_monitoring}. Figure~\ref{fig:framework} illustrates the overall architecture of \nm, comprising four key phases:
\textcircled{\raisebox{-0.9pt}{1}} \nm identifies LLM training jobs from tens of thousands of GPUs based on spatial communication patterns (§~\ref{sec:method_phase1}).
\textcircled{\raisebox{-0.9pt}{2}} For each training job, \nm robustly identify parallelism strategies (\ie classify DP and PP pairs) according to their distinct characteristics (§~\ref{sec:method_phase2}).
\textcircled{\raisebox{-0.9pt}{3}} For each GPU rank, \nm reconstructs the training timeline by analyzing temporal communication patterns (§~\ref{sec:method_phase3}).
\textcircled{\raisebox{-0.9pt}{4}} By performing multi-dimensional anomaly detection of reconstructed training timelines, \nm effectively detects performance issues in LLM training, enabling timely alerts and mitigation (§~\ref{sec:method_phase4}).

\begin{algorithm}[t]
\small
\caption{LLM Training Jobs Recognition}
\label{algo:phase1}
\SetAlgoLined
\KwIn{RoCE Network Flows: $\mathcal{F}=\{f_1, f_2, \cdots\}$; List of GPUs: $\mathcal{X}=\{x_1, x_2, \cdots \}$; Physical Topology: $T$}
\KwOut{Recognized LLM task jobs: $\mathcal{C}=\{C_1, C_2, \cdots \}$;}

$U \leftarrow$ Disjoint-set data structure

// (1) Search neighbors and build cross-machine clusters

\For{each network flow $f_i \in \mathcal{F}$ }{ \label{algo:step1_begin}
    srcAddr, dstAddr $\leftarrow$ $f_i$

    \If{$U$.findSet(srcAddr) != $U$.findSet(dstAddr)} {
            $U$.unionSet(srcAddr, dstAddr) // merge two GPUs\label{algo:step1_end}
        }
}
$\mathcal{CR} \leftarrow U$.getAllSets() // cross-machine clusters \\

// (2) Build task-level clusters based on physical topology $T$

\For{each cross-machine clusters $c_i \in \mathcal{CR}$ }{ \label{algo:step2_begin}
    get all server list $s_i$ of $c_i$ from physical topology
}
\For{each pair $c_i$ and $c_j \in \mathcal{CR}$} {
    \If{Jaccard Similarity ($s_i$, $s_j$) = 1} {
        merge $c_i$ and $c_j$
    }
}\label{algo:step2_end}
\Return $\mathcal{C} \leftarrow$ getMergedSet() // job-level clusters
\end{algorithm}

\subsection{LLM Training Jobs Recognition}
\label{sec:method_phase1}

From the perspective of LLM training service providers, training platforms function as black boxes.
Service providers have visibility only into the GPU machines rented by each tenant.
However, tenants can deploy multiple LLM training jobs on their rented GPU machines. Due to privacy considerations, the platform’s site reliability engineers (SREs) lack detailed knowledge of the configurations of these jobs. For instance, they are unaware of the number of machines utilized or the specific machines involved in each job.

This black-box nature hinders SREs from efficiently and effectively identifying and diagnosing potential issues within LLM training platforms. 
Therefore, the first step in transforming this black-box model into a white-box one is to recognize LLM training jobs among the vast number of GPUs within the platform.
This recognition is feasible based on observations of LLM training communication patterns: communications between ranks are static, meaning that \textit{only GPUs within the same LLM training job will communicate and generate corresponding network flows.}

Building on this insight, we propose an efficient algorithm to identify LLM training jobs based on network flows over a short time window (\eg one minute), as outlined in Alg.~\ref{algo:phase1}.
Specifically, we employ a disjoint-set data structure to maintain clusters of GPUs.
For each network flow, we extract communication GPU pairs and merge the clusters to which they belong, as these GPUs are engaged in the same training job (lines~\ref{algo:step1_begin} – \ref{algo:step1_end}).
Ultimately, each cluster represents a group of GPUs engaged in cross-machine communication within the same LLM training job, referred to as \emph{cross-machine} clusters.

Nevertheless, as discussed in Sec.~\ref{background:training}, tensor parallelism (TP) involves intra-machine communications that are confined to a single machine.
Such traffic remains unobservable to network switches, preventing the correlation of GPU ranks within the same machine using network flow data alone.
To address this, we incorporate the physical topology of the cluster, information available to platform providers, to further merge cross-machine clusters. 
Specifically, for each cross-machine cluster, we record the set of physical machine addresses associated with it and merge clusters that share the same set (lines~\ref{algo:step2_begin} – \ref{algo:step2_end}).
This process yields complete job-level clusters, each encompassing all GPUs involved in the same training job.

\subsection{Parallelism Strategies Identification}
\label{sec:method_phase2}

After identifying LLM training jobs, it remains essential to analyze their parallel communication strategies to accurately reconstruct the training process.
Our primary insight is that \textit{different cross-machine communication strategies, \ie data parallelism (DP) and pipeline parallelism (PP), exhibit distinct characteristics}, enabling the differentiation of their communication types. 
Specifically, point-to-point PP communications (\eg send and receive) occur within each micro-batch during a single training step, overlaps with computation, and transmits gradients and loss values.
Consequently, its communication volume is small, and the corresponding network flows for each PP communication pair exhibit consistent sizes.
In contrast, DP involves collective communication at the conclusion of each training step to synchronize gradients and model parameters.
Given its large communication volume, DP communication typically divides into multiple network flows with varying sizes.
Additionally, practical challenges such as data noise, arising from packet loss, retransmission, or incomplete flow collection, necessitate a robust algorithm to ensure accurate identification of communication types.

Based on these distinct characteristics, we design a robust algorithm, detailed in Alg.~\ref{algo:phase2}, to precisely classify communication types within individual LLM training jobs.
For each GPU communication pair, the algorithm initially categorizes network flows according to their respective training steps (line \ref{algo:phase2_step_1_begin} - \ref{algo:phase2_step_1_end}).
This categorization leverages the observation that \textit{intervals between communication flows within the same step are significantly shorter than those between adjacent steps}.
Accordingly, we compute the time intervals ($\Delta t$) between sequential network flows and apply Bayesian Online Change-point Detection (BOCD)~\cite{agudelo2020bayesian} to automatically determine step boundaries.
BOCD is an efficient, linear-time algorithm to detect change-points in dynamic sequences.
Specifically, BOCD defines a run-length $r_t$ at each timestamp $t$ as follows:
\begin{equation}
\small
r_t =
\begin{cases}
0, & \text{if a change-point occurs at } t\\
r_{t - 1}+1, & \text{otherwise}
\end{cases}.
\end{equation}
BOCD applies Bayesian inference to compute the likelihood that $r_t = 0$ (a change-point at $t$) for every timestamp.
If this likelihood exceeds a predefined threshold (set to 0.95 in our implementation), timestamp $t$ is reported as a change-point.

After dividing the flows into training steps, we count the distinct flow sizes, denoted as $N_k$, within the $k$-th step for each pair.
The mode of these values is computed to mitigate interference from noisy data.
If Mode($N_k$) is one, the communication pair is classified as PP; otherwise, it is classified as DP (line \ref{algo:phase2_step_2_begin} - \ref{algo:phase2_step_2_end}).
Our empirical results indicate that this approach accurately classifies all PP pairs, although certain DP pairs may initially be misclassified as PP.
To address this, we utilize the transitive property inherent in DP communications for further refinement (line \ref{algo:phase2_step_3_begin} - \ref{algo:phase2_step_3_end}).
That is, if $(u, v)$ is identified as a DP pair and $(v, r)$ is also a DP pair, then $(u, r)$ must be a DP pair.
We apply a depth-first search (DFS) algorithm to identify all connected components within the DP communication graph $\mathcal{G}$, subsequently refining the classification accordingly.
These methodological steps enhance the robustness and accuracy of our communication type identification process.

\begin{algorithm}[t]
\small
\caption{Communication Type Identification}
\label{algo:phase2}
\SetAlgoLined
\KwIn{RoCE Network Flows: $\mathcal{F}=\{f_1, f_2, \cdots\}$; Comm. pairs within a job: $\mathcal{P}=\{(u_1, v_1), \cdots (u_n, v_n)\}$}
\KwOut{Identified communication types: $\mathcal{T}=\{t_1, \cdots, t_n\}$;}

$\mathcal{G} \leftarrow$ unordered graph of all GPUs within the job \\


\For{each pair $(u, v)$ in $\mathcal{P}$} {
// (1) Calculate communication interval\label{algo:phase2_step_1_begin} \\
    extract related network flows: $\mathcal{F}_{(u,v)}=\{f_1, \cdots, f_t\}$ \\
    compute comm. interval: $\Delta t_i = f_{i+1}(time) - f_i(time)$ \\
// (2) Step division \\
    employ Bayesian CPD: $\mathcal{F}_{(u,v)} = \{F_{step_1}, F_{step_2}, \cdots \}$ \label{algo:phase2_step_1_end} \\
// (3) Type distinction \\
    \For{each $step_k$}  { \label{algo:phase2_step_2_begin}
        count the distinct number of flow sizes $N_k$ in $F_{step_k}$ \\
    }
    type $t(u, v)$ $\leftarrow$ PP if Mode($N_k$) = 1 else DP \\
    add edge $(u, v)$ to $\mathcal{G}$ if $t(u, v)$ is DP \\ \label{algo:phase2_step_2_end}
}
// (4) Noise refinement \\
DFS to find all connected component $CC_i$ in $\mathcal{G}$ \label{algo:phase2_step_3_begin} \\
\For{all pairs $(u, v)$ in all $CC_i$} {
    refine $t(u, v)$ to DP if it is PP
} \label{algo:phase2_step_3_end}
\Return $\mathcal{T}=\{t_1, \cdots, t_n\}$ // final communication types
\end{algorithm}

\subsection{Training Timeline Reconstruction}
\label{sec:method_phase3}

Building upon the inferred LLM training jobs and parallelism strategies, \nm further reconstruct the training timeline for each GPU.
Various optimization techniques, such as DeepSpeed-ZeRO~\cite{rasley2020deepspeed}, have been proposed and employed to overlap communication and computation during the training process.
Although platform providers typically lack access to detailed optimizations, we identify a consistent temporal pattern: \textit{each training step concludes with a segment of DP collective communication traffic}, as disccused in Sec.~\ref{background:training}.

Leveraging this observation, \nm performs a training timeline reconstruction for each GPU in an online manner.
Specifically, analogous to the step division described in Sec.~\ref{sec:method_phase2}, we apply the BOCD algorithm to identify change-points in intervals between DP communication flows for each rank.
This method facilitates partitioning DP communication flows into discrete training steps, with the conclusion of DP flows marking the end of each training step.
Through this method of step boundary delineation, we reconstruct a detailed chronological timeline of both PP and DP communication events for each GPU rank.
Fig.~\ref{fig:framework} demonstrates an example of such a reconstructed training timeline, where the intervals between communication events are approximated as computing operations.
This reconstructed timeline visualization is also incorporated into our SRE platform, offering a comprehensive view of each GPU's training procedure, thereby supporting further analysis and performance diagnosis.

\subsection{Performance Diagnosis}
\label{sec:method_phase4}

\begin{figure}[t]
    \centering
    \includegraphics[width=\columnwidth]{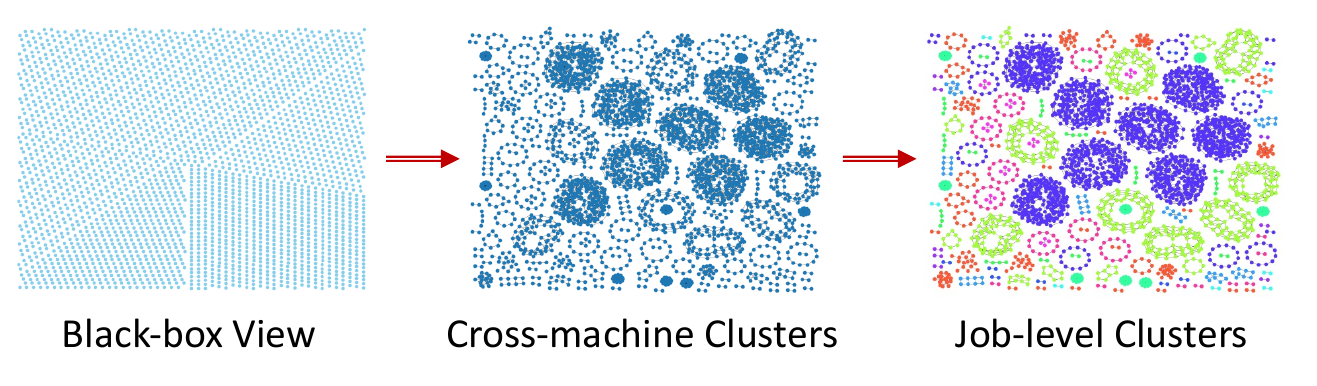}
    \caption{LLM training job recognition procedure of \nm in training clusters with 2,880 GPUs in \platform.}
    \label{fig:job_recognition}
\end{figure}

The accurate reconstruction of the LLM training timeline facilitates various performance diagnosis methodologies, akin to those employed in existing intrusive profiling approaches~\cite{wu2024falcon, HTA,cui2025xputimer}.
Specifically, \nm employs multi-dimensional performance degradation diagnosis as follows:
\begin{itemize}[leftmargin=*, topsep=0pt, parsep=0pt]
\item \textbf{Cross-step Diagnosis.}
Under normal conditions, the duration of each training step remains consistent and stable. Deviations or increases in step duration typically signify performance anomalies such as computational delays or slow network communication. Therefore, \nm continuously monitors the duration of reconstructed training steps to detect potential performance issues.
\item \textbf{Cross-group Diagnosis.}
To further identify potential network-related problems, such as network congestion, \nm adopts cross-group detection. As observed in previous studies~\cite{wu2024falcon}, most network issues occur within collective DP operations (\eg AllReduce) because their communication volume substantially exceeds that of PP operations. Consequently, \nm primarily focuses on DP communication groups and their respective communication efficiencies. Ideally, DP communication durations across different DP groups should exhibit minimal variance. \nm thus records and analyzes DP communication durations within each step to identify anomalous delays indicating potential network issues.
\item \textbf{Switch-level Diagnosis.}
Leveraging switch-level information from network flows, \nm provides fine-granular analysis of network anomalies. Specifically, \nm aggregates network flows through individual switches.
On one hand, \nm counts distinct DP flows based on identified communication types detailed in Sec.~\ref{sec:method_phase2}. Given that DP flows require substantial network bandwidth, exceeding concurrent DP communication limits can lead to congestion within a switch. \nm effectively identifies such configuration-induced network issues, promptly alerting SREs and tenants.
On the other hand, \nm computes average DP communication bandwidth per switch within each step and performs degradation analysis across switches to identify potential network bottlenecks.
\end{itemize}
For anomaly detection used in different levels, \nm selects the straightforward yet effective k-$\sigma$ rules~\cite{son2016anomaly}, as it does not require specific thresholds.
In detail, it classifies data points exceeding $\bar{l}$ + k$\sigma$ as outliers, where $\sigma = \frac{1}{n}\sum_{i=1}^{n} (l_i - \bar{l})$ and k is set to 3.
Through these multi-dimensional performance degradation detection mechanisms, \nm provides comprehensive online monitoring of LLM training jobs, enabling efficient identification and mitigation of performance issues.

\section{Evaluation and Deployed Experiences}

\nm has been deployed in our multi-tenant LLM training \platform serving hundreds of internal users and partner companies for six months.
Due to tenant privacy constraints, we are unable to access and share tenants’ job configurations for large-scale experiments.
Consequently, we validate its effectiveness through case studies based on training jobs from internal tenants and share our deployment experiences to benefit both practitioners and researchers in this field.

\begin{table}[tbp]
    \centering
    \caption{Accuracy in identifying parallelism strategies via flows of varying-length periods for jobs with 1024 GPUs.}
    \label{tab:parallism_identification}
    \resizebox{\linewidth}{!}{%
        \begin{NiceTabular}{l||cccc}
        \CodeBefore
        \Body
            \toprule
                \rowcolor{gray!15} Methods &  1 min Acc. & 3 min Acc. & 5 min Acc. & 10min Acc.\\
                \midrule
                \nm w/o refinement & 96.00\% & 97.93\% & 98.03\%  & 99.61\% \\
                \nm & 100\% &  100\% & 100\% & 100\% \\
            \bottomrule
             \end{NiceTabular}%
     }
\end{table}

\subsection{Effectiveness of LLM Training Jobs Recognition}

To validate the accuracy of \nm in recognizing LLM training jobs under a black-box view through network flow analysis, we conduct experiments on an internal training cluster consisting of 2,880 GPUs within \platform.
Specifically, \nm analyzes the collected network flow over a one-minute time window and applies the training job recognition algorithm described in §~\ref{sec:method_phase1}.
As illustrated in Fig.~\ref{fig:job_recognition}, the left panel presents the black-box view of the training cluster, where SREs lack visibility into the relationships among the large number of GPUs.
Leveraging the spatial patterns of LLM training communication, \nm effectively identifies cross-machine GPU clusters, where each cluster comprises GPUs that engage in cross-machine communication within a single training job.
Furthermore, by incorporating the physical topology of the cluster, \nm accurately reconstructs complete job-level clusters, as depicted in the right panel of Fig.~\ref{fig:job_recognition}, where clusters of the same color correspond to the same LLM training job.
Finally, \nm successfully identifies 19 training jobs across the 2,880 GPUs, with manual verification conducted in collaboration with internal tenants, demonstrating its accuracy and effectiveness.

\subsection{Effectiveness of Parallelism Strategies Identification}

To evaluate the accuracy of \nm in identifying parallelism strategies, \ie the inferred types (DP and PP) of communication pairs within each training job, we select five LLM training jobs with 1024 GPUs from our internal tenants, for which we have access to ground-truth parallelism configurations.
These jobs involve diverse trained LLMs (\eg the LLaMA family~\cite{LlamaFamily}) and various parallelism optimization techniques (\eg DeepSpeed-ZeRO~\cite{rasley2020deepspeed}).

Specifically, we analyze network flow data over time windows of varying lengths to apply \nm and compute accuracy as the ratio of correctly classified communication pairs to the total number of communication pairs.
The average results, presented in Tab.~\ref{tab:parallism_identification}, lead to the following observations:
(1) Without the noise refinement process introduced in §~\ref{sec:method_phase2}, errors may occur in parallelism strategy classification, where some DP pairs are misclassified as PP pairs. This issue becomes more pronounced when network flow data is limited to shorter durations. For example, without refinement, the  accuracy of \nm  drops from 99.61\% to 96.00\% as the time window length decreases from 10 minutes to just 1 minute.
(2) Incorporating the noise refinement process allows \nm to consistently achieve 100\% accuracy across all training jobs, even when using network flow data from only a 1-minute period. These results demonstrate the effectiveness and robustness of \nm in accurately identifying LLM training parallelism strategies.

\subsection{Effectiveness of Training Timeline Reconstruction}

\begin{figure}[t]
    \centering
    \includegraphics[width=\columnwidth]{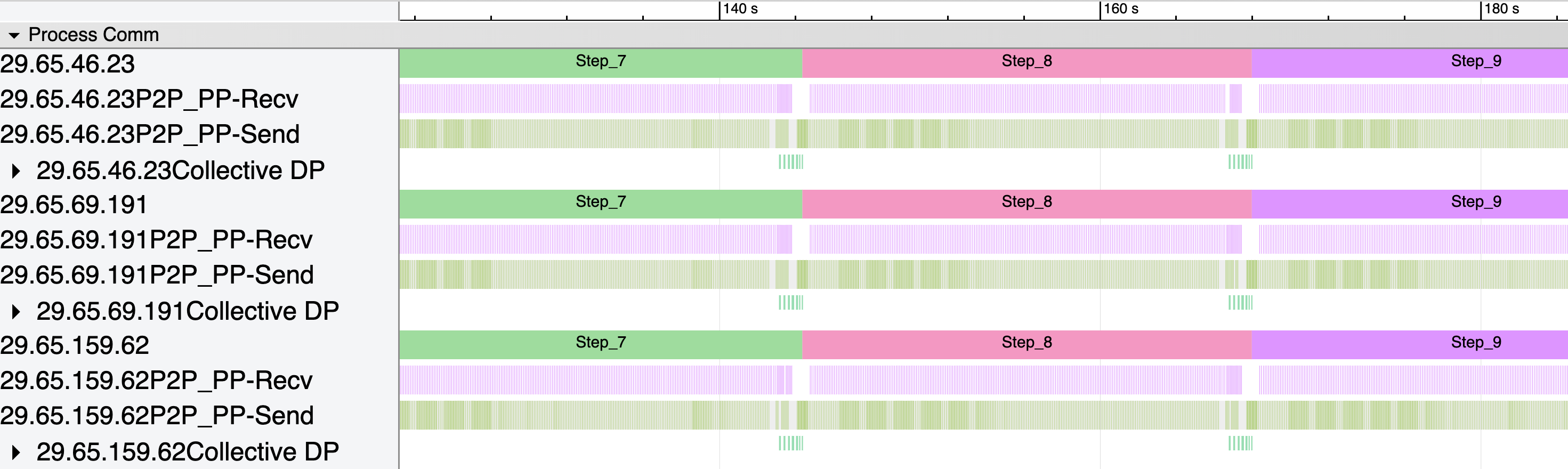}
    \caption{An example of reconstructed training timeline.}
    \label{fig:timeline_reconstruction}
\end{figure}

To quantify the accuracy of the reconstructed training timeline, we create a training job using 1,024 GPUs in \platform.
Instead of network flows, we also collect the training process logs and profiling results from the PyTorch Profiler~\cite{torhcprofiler}.
By comparing the training timeline reconstructed by \nm with these reference data, we found that the reconstruction error of \nm remained within 0.3\%, demonstrating its high precision.
Figure~\ref{fig:timeline_reconstruction} presents an example visualization of the reconstructed training timeline, illustrating the chronological interleaving of communication operations for each GPU rank, including PP (\eg send and receive) and DP operations (\eg all-reduce).
This visualization provides SREs with a comprehensive overview of the training process, facilitating in-depth inspection and analysis of specific LLM training jobs.

\subsection{Effectiveness of Performance Diagnosis}

\begin{figure}[t]
    \centering
    \includegraphics[width=0.85\columnwidth]{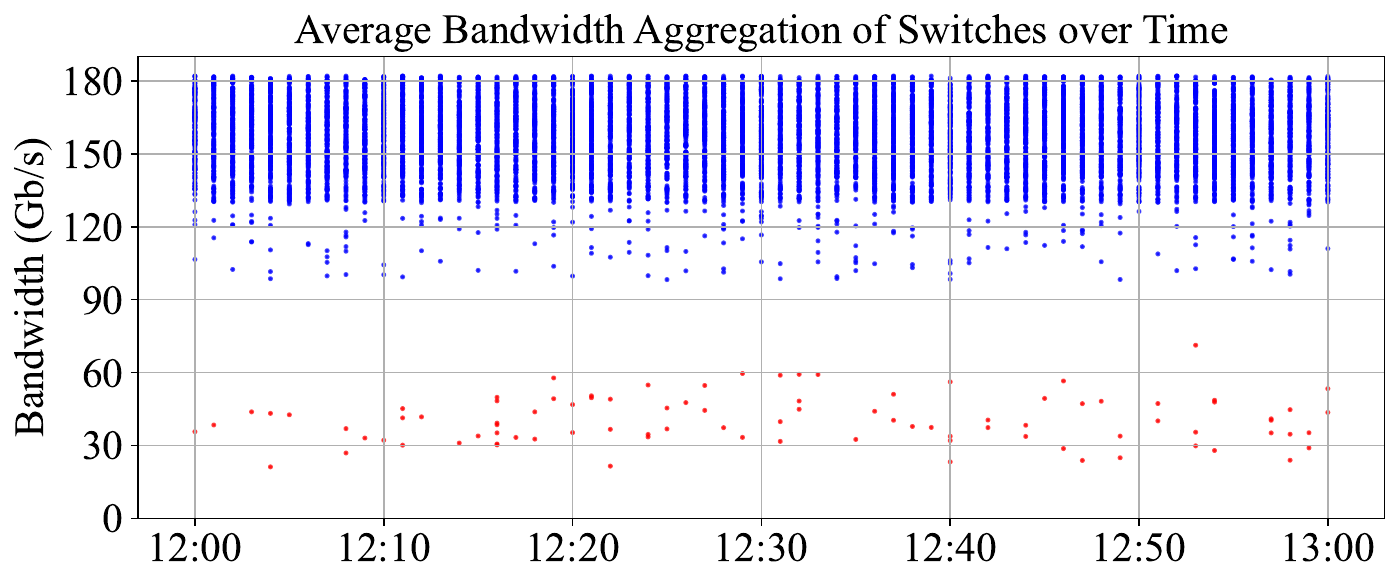}
    \caption{Switch-level diagnosis of \nm in \platform using one-hour aggregated average bandwidths.}
    \label{fig:bandwidth}
\end{figure}

Based on the precise reconstruction of LLM training timelines, \nm enables effective cross-step and cross-group diagnosis by detecting delayed steps or DP groups, as detailed in §~\ref{sec:method_phase4}.
Since its deployment in \platform, \nm has autonomously identified and alerted a substantial number of fail-slow cases, the majority of which have been manually confirmed by SREs.
Moreover, \nm is capable of providing deeper insights into network issues through switch-level diagnostics.
Figure~\ref{fig:bandwidth} illustrates a case of the average DP flow bandwidth of each switch within our training cluster over a one-hour window.
While typical average bandwidth ranges between 100 Gb/s and 180 Gb/s, abnormal network behavior was observed during this specific period, \ie a marked degradation in the bandwidth of a subset of switches to approximately 30-60 Gb/s. Consequently, \nm triggered alerts notifying SREs of these anomalous switches, prompting further investigation into potential network congestion issues that could affect the LLM training process.

\section{Related Work}

Monitoring and profiling LLM training jobs are critical for ensuring optimal performance and diagnosing issues within the training process.
To this end, several studies and profiling tools have been conducted.
For instance, PyTorch provides Torch Profiler~\cite{torhcprofiler}, which exhaustively traces CUDA events to monitor the training progression.
Meta's Strobelight~\cite{strobelight} leverages eBPF~\cite{ebpf} to hook GPU events and measure the duration of various operations.
Furthermore, XPUTimer~\cite{cui2025xputimer} employs a tracing daemon attached to each training process to measure latencies in critical code segments.
However, these existing methods rely on specific frameworks or tenant-configured settings, which is impractical from the perspective of the training platform providers.
Additionally, these profilers inevitably introduce overhead and impact training efficiency.
In contrast, \nm introduces an innovative approach that utilizes collected network flows to precisely reconstruct the training timeline, operating independently from the training process.
This method achieves non-intrusive profiling with near-zero performance overhead, making it particularly suitable for large-scale LLM training platforms.

\section{Generalization Discussion}

This study presents \nm, our proposed and deployed closed-box monitoring solution, integrated into our large-scale LLM training \platform.
We believe that \nm can be adopted across diverse LLM training platforms, providing non-intrusive and continuous performance diagnosis capabilities.
On the one hand, our \platform utilizes widely adopted technologies and maintains architectural similarities with other leading systems~\cite{jeon2019analysis,jiang2024megascale,kokolis2024revisiting,dongevolution}, including high-speed inter-connected networks.
Furthermore, training jobs on \platform span various widely utilized LLMs (\eg LLaMA~\cite{LlamaFamily}), along with prevalent training frameworks and optimization strategies (\eg PyTorch~\cite{pytorch} and DeepSpeed~\cite{deepspeed}).
Consequently, the spatial and temporal patterns, alongside the communication characteristics utilized by \nm, exhibit generazability to other training platforms and jobs, highlighting its values to enhance the reliability of large-scale LLM training systems.
\section{Conclusion}

In this paper, we introduce \nm, the first black-box performance diagnostic framework designed specifically for large-scale LLM training platforms based on network analysis. Leveraging the distinct features inherent in LLM training communication, \nm utilizes network flows to progressively identify individual training jobs, recognize their underlying parallelism strategies, and reconstruct precise training timelines. This approach enables non-intrusive and continuous performance diagnosis for LLM training jobs with minimal performance impact.
The evaluation and deployment of \nm on our production \platform validate its effectiveness and practicality. Given the diverse application scenarios and the generalizable observations utilized by \nm, we believe that it can be easily applied to other platforms, offering valuable insights to both researchers and practitioners in this field.

\section*{Acknowledgement}

This work was supported by the Research Grants Council of the Hong Kong Special Administrative Region, China (No. SRFS2425-4S03 of the Senior Research Fellow Scheme and No. CUHK 14209124 of the General Research Fund).

\balance
\normalem
\bibliographystyle{IEEEtran}
\bibliography{DSN25}

\begin{thebibliography}{10}
\providecommand{\url}[1]{#1}
\csname url@samestyle\endcsname
\providecommand{\newblock}{\relax}
\providecommand{\bibinfo}[2]{#2}
\providecommand{\BIBentrySTDinterwordspacing}{\spaceskip=0pt\relax}
\providecommand{\BIBentryALTinterwordstretchfactor}{4}
\providecommand{\BIBentryALTinterwordspacing}{\spaceskip=\fontdimen2\font plus
\BIBentryALTinterwordstretchfactor\fontdimen3\font minus \fontdimen4\font\relax}
\providecommand{\BIBforeignlanguage}[2]{{%
\expandafter\ifx\csname l@#1\endcsname\relax
\typeout{** WARNING: IEEEtran.bst: No hyphenation pattern has been}%
\typeout{** loaded for the language `#1'. Using the pattern for}%
\typeout{** the default language instead.}%
\else
\language=\csname l@#1\endcsname
\fi
#2}}
\providecommand{\BIBdecl}{\relax}
\BIBdecl

\bibitem{jin2024comprehensive}
H.~Jin, Y.~Zhang, D.~Meng, J.~Wang, and J.~Tan, ``A comprehensive survey on process-oriented automatic text summarization with exploration of llm-based methods,'' \emph{arXiv preprint arXiv:2403.02901}, 2024.

\bibitem{liu2023scalable}
J.~Liu, J.~Huang, Y.~Huo, Z.~Jiang, J.~Gu, Z.~Chen, C.~Feng, M.~Yan, and M.~R. Lyu, ``Scalable and adaptive log-based anomaly detection with expert in the loop,'' \emph{arXiv preprint arXiv:2306.05032}, 2023.

\bibitem{li2024exploring}
Y.~Li, Y.~Huo, Z.~Jiang, R.~Zhong, P.~He, Y.~Su, L.~C. Briand, and M.~R. Lyu, ``Exploring the effectiveness of llms in automated logging statement generation: An empirical study,'' \emph{IEEE Transactions on Software Engineering}, 2024.

\bibitem{jiang2024lilac}
Z.~Jiang, J.~Liu, Z.~Chen, Y.~Li, J.~Huang, Y.~Huo, P.~He, J.~Gu, and M.~R. Lyu, ``Lilac: Log parsing using llms with adaptive parsing cache,'' \emph{Proceedings of the ACM on Software Engineering}, vol.~1, no. FSE, pp. 137--160, 2024.

\bibitem{li2024go}
Y.~Li, Y.~Huo, R.~Zhong, Z.~Jiang, J.~Liu, J.~Huang, J.~Gu, P.~He, and M.~R. Lyu, ``Go static: Contextualized logging statement generation,'' \emph{Proceedings of the ACM on Software Engineering}, vol.~1, no. FSE, pp. 609--630, 2024.

\bibitem{li2025coca}
Y.~Li, Y.~Wu, J.~Liu, Z.~Jiang, Z.~Chen, G.~Yu, and M.~Lyu, ``Coca: Generative root cause analysis for distributed systems with code knowledge,'' in \emph{2025 IEEE/ACM 47th International Conference on Software Engineering (ICSE)}.\hskip 1em plus 0.5em minus 0.4em\relax IEEE Computer Society, 2025, pp. 770--770.

\bibitem{kaplan2020scaling}
J.~Kaplan, S.~McCandlish, T.~Henighan, T.~B. Brown, B.~Chess, R.~Child, S.~Gray, A.~Radford, J.~Wu, and D.~Amodei, ``Scaling laws for neural language models,'' \emph{arXiv preprint arXiv:2001.08361}, 2020.

\bibitem{guo2025deepseek}
D.~Guo, D.~Yang, H.~Zhang, J.~Song, R.~Zhang, R.~Xu, Q.~Zhu, S.~Ma, P.~Wang, X.~Bi \emph{et~al.}, ``Deepseek-r1: Incentivizing reasoning capability in llms via reinforcement learning,'' \emph{arXiv preprint arXiv:2501.12948}, 2025.

\bibitem{he2023unicron}
T.~He, X.~Li, Z.~Wang, K.~Qian, J.~Xu, W.~Yu, and J.~Zhou, ``Unicron: Economizing self-healing llm training at scale,'' \emph{arXiv preprint arXiv:2401.00134}, 2023.

\bibitem{DBLP:conf/sc/AnBCCDDDDGGGGFH24}
W.~An, X.~Bi, G.~Chen, S.~Chen, C.~Deng, H.~Ding, K.~Dong, Q.~Du, W.~Gao, K.~Guan \emph{et~al.}, ``Fire-flyer ai-hpc: A cost-effective software-hardware co-design for deep learning,'' in \emph{SC24: International Conference for High Performance Computing, Networking, Storage and Analysis}.\hskip 1em plus 0.5em minus 0.4em\relax IEEE, 2024, pp. 1--23.

\bibitem{dubey2024llama}
A.~Dubey, A.~Jauhri, A.~Pandey, A.~Kadian, A.~Al-Dahle, A.~Letman, A.~Mathur, A.~Schelten, A.~Yang, A.~Fan \emph{et~al.}, ``The llama 3 herd of models,'' \emph{arXiv preprint arXiv:2407.21783}, 2024.

\bibitem{amazon-ai}
\BIBentryALTinterwordspacing
``Amazon sagemaker,'' 2024. [Online]. Available: \url{https://aws.amazon.com/sagemaker/}
\BIBentrySTDinterwordspacing

\bibitem{googlevertex}
\BIBentryALTinterwordspacing
``Google vertex ai,'' 2024. [Online]. Available: \url{https://console.cloud.google.com/vertex-ai?hl=en&inv=1&invt=Abkx0g&project=fine-effect-362306}
\BIBentrySTDinterwordspacing

\bibitem{dongevolution}
J.~Dong, K.~Qian, P.~Zhang, Z.~Zheng, L.~Chen, F.~Feng, Y.~Zhu, G.~Lu, Z.~Ren, X.~Li \emph{et~al.}, ``Evolution of aegis: Fault diagnosis for ai model training cloud service in production (experience track).''

\bibitem{liu2023prism}
J.~Liu, Z.~Jiang, J.~Gu, J.~Huang, Z.~Chen, C.~Feng, Z.~Yang, Y.~Yang, and M.~R. Lyu, ``Prism: Revealing hidden functional clusters from massive instances in cloud systems,'' in \emph{2023 38th IEEE/ACM International Conference on Automated Software Engineering (ASE)}.\hskip 1em plus 0.5em minus 0.4em\relax IEEE, 2023, pp. 268--280.

\bibitem{huang2024faultprofit}
J.~Huang, J.~Liu, Z.~Chen, Z.~Jiang, Y.~Li, J.~Gu, C.~Feng, Z.~Yang, Y.~Yang, and M.~R. Lyu, ``Faultprofit: Hierarchical fault profiling of incident tickets in large-scale cloud systems,'' in \emph{Proceedings of the 46th International Conference on Software Engineering: Software Engineering in Practice}, 2024, pp. 392--404.

\bibitem{wu2024falcon}
T.~Wu, W.~Wang, Y.~Yu, S.~Yang, W.~Wu, Q.~Duan, G.~Yang, J.~Wang, L.~Qu, and L.~Zhang, ``Falcon: Pinpointing and mitigating stragglers for large-scale hybrid-parallel training,'' \emph{arXiv preprint arXiv:2410.12588}, 2024.

\bibitem{jiang2024megascale}
Z.~Jiang, H.~Lin, Y.~Zhong, Q.~Huang, Y.~Chen, Z.~Zhang, Y.~Peng, X.~Li, C.~Xie, S.~Nong \emph{et~al.}, ``Megascale: Scaling large language model training to more than 10,000 gpus,'' in \emph{21st USENIX Symposium on Networked Systems Design and Implementation (NSDI 24)}, 2024, pp. 745--760.

\bibitem{kokolis2024revisiting}
A.~Kokolis, M.~Kuchnik, J.~Hoffman, A.~Kumar, P.~Malani, F.~Ma, Z.~DeVito, S.~Sengupta, K.~Saladi, and C.-J. Wu, ``Revisiting reliability in large-scale machine learning research clusters,'' \emph{arXiv preprint arXiv:2410.21680}, 2024.

\bibitem{jiang2025l4}
Z.~Jiang, J.~Huang, Z.~Chen, Y.~Li, G.~Yu, C.~Feng, Y.~Yang, Z.~Yang, and M.~R. Lyu, ``L4: Diagnosing large-scale llm training failures via automated log analysis,'' \emph{arXiv preprint arXiv:2503.20263}, 2025.

\bibitem{jin2024communication}
H.~Jin, ``Communication-efficient distributed llm training with adaptive traffic control,'' in \emph{2024 IEEE International Conference on Big Data (BigData)}.\hskip 1em plus 0.5em minus 0.4em\relax IEEE, 2024, pp. 8685--8687.

\bibitem{feng2024optimus}
W.~Feng, Y.~Chen, S.~Wang, Y.~Peng, H.~Lin, and M.~Yu, ``Optimus: Accelerating large-scale multi-modal llm training by bubble exploitation,'' \emph{arXiv preprint arXiv:2408.03505}, 2024.

\bibitem{patel2024accelerating}
A.~Patel, D.~Biswas, J.~Kundu, Y.~Ban, N.~Pantano, A.~Mallik, J.~Ryckaert, and J.~Myers, ``Accelerating large language model training with in-package optical links for scale-out systems,'' in \emph{2024 IEEE Computer Society Annual Symposium on VLSI (ISVLSI)}.\hskip 1em plus 0.5em minus 0.4em\relax IEEE, 2024, pp. 118--123.

\bibitem{cui2025xputimer}
W.~Cui, J.~Zhang, H.~Zhao, C.~Liu, W.~Zhang, J.~Sha, Q.~Chen, B.~He, and M.~Guo, ``Xputimer: Anomaly diagnostics for divergent llm training in gpu clusters of thousand-plus scale,'' \emph{arXiv preprint arXiv:2502.05413}, 2025.

\bibitem{HTA}
\BIBentryALTinterwordspacing
``Holistic trace analysis, pytorch,'' 2024. [Online]. Available: \url{https://pytorch.org/tutorials/beginner/hta_intro_tutorial.html}
\BIBentrySTDinterwordspacing

\bibitem{strobelight}
\BIBentryALTinterwordspacing
``Meta strobelight,'' 2025. [Online]. Available: \url{https://engineering.fb.com/2025/01/21/production-engineering/strobelight-a-profiling-service-built-on-open-source-technology/}
\BIBentrySTDinterwordspacing

\bibitem{ERSPAN}
\BIBentryALTinterwordspacing
``Erspan,'' 2024. [Online]. Available: \url{https://netseccloud.com/understanding-span-ports}
\BIBentrySTDinterwordspacing

\bibitem{kaur2013rdma}
G.~Kaur and M.~Bala, ``Rdma over converged ethernet: A review,'' \emph{International Journal of Advances in Engineering \& Technology}, vol.~6, no.~4, p. 1890, 2013.

\bibitem{shoeybi2019megatron}
M.~Shoeybi, M.~Patwary, R.~Puri, P.~LeGresley, J.~Casper, and B.~Catanzaro, ``Megatron-lm: Training multi-billion parameter language models using model parallelism,'' \emph{arXiv preprint arXiv:1909.08053}, 2019.

\bibitem{rajasekaran2024cassini}
S.~Rajasekaran, M.~Ghobadi, and A.~Akella, ``$\{$CASSINI$\}$:$\{$Network-Aware$\}$ job scheduling in machine learning clusters,'' in \emph{21st USENIX Symposium on Networked Systems Design and Implementation (NSDI 24)}, 2024, pp. 1403--1420.

\bibitem{zheng2022alpa}
L.~Zheng, Z.~Li, H.~Zhang, Y.~Zhuang, Z.~Chen, Y.~Huang, Y.~Wang, Y.~Xu, D.~Zhuo, E.~P. Xing \emph{et~al.}, ``Alpa: Automating inter-and $\{$Intra-Operator$\}$ parallelism for distributed deep learning,'' in \emph{16th USENIX Symposium on Operating Systems Design and Implementation (OSDI 22)}, 2022, pp. 559--578.

\bibitem{gangidi2024rdma}
A.~Gangidi, R.~Miao, S.~Zheng, S.~J. Bondu, G.~Goes, H.~Morsy, R.~Puri, M.~Riftadi, A.~J. Shetty, J.~Yang \emph{et~al.}, ``Rdma over ethernet for distributed training at meta scale,'' in \emph{Proceedings of the ACM SIGCOMM 2024 Conference}, 2024, pp. 57--70.

\bibitem{zhang2023opt}
S.~Zhang, S.~Roller, N.~Goyal, M.~Artetxe, M.~Chen, S.~Chen, C.~Dewan, M.~Diab, X.~Li, X.~V. Lin \emph{et~al.}, ``Opt: Open pre-trained transformer language models, 2022,'' \emph{URL https://arxiv. org/abs/2205.01068}, vol.~3, pp. 19--0, 2023.

\bibitem{agudelo2020bayesian}
D.~Agudelo-Espa{\~n}a, S.~Gomez-Gonzalez, S.~Bauer, B.~Sch{\"o}lkopf, and J.~Peters, ``Bayesian online prediction of change points,'' in \emph{Conference on uncertainty in artificial intelligence}.\hskip 1em plus 0.5em minus 0.4em\relax PMLR, 2020, pp. 320--329.

\bibitem{rasley2020deepspeed}
J.~Rasley, S.~Rajbhandari, O.~Ruwase, and Y.~He, ``Deepspeed: System optimizations enable training deep learning models with over 100 billion parameters,'' in \emph{Proceedings of the 26th ACM SIGKDD international conference on knowledge discovery \& data mining}, 2020, pp. 3505--3506.

\bibitem{son2016anomaly}
S.~Son, M.-S. Gil, Y.-S. Moon, and H.-S. Won, ``Anomaly detection of hadoop log data using moving average and 3-sigma,'' \emph{KIPS Transactions on Software and Data Engineering}, vol.~5, no.~6, pp. 283--288, 2016.

\bibitem{LlamaFamily}
\BIBentryALTinterwordspacing
``The llama family,'' 2025. [Online]. Available: \url{https://huggingface.co/meta-llama}
\BIBentrySTDinterwordspacing

\bibitem{torhcprofiler}
\BIBentryALTinterwordspacing
``Torch profiler,'' 2025. [Online]. Available: \url{https://pytorch.org/docs/stable/profiler.html}
\BIBentrySTDinterwordspacing

\bibitem{ebpf}
\BIBentryALTinterwordspacing
``ebpf documentary,'' 2025. [Online]. Available: \url{https://ebpf.io}
\BIBentrySTDinterwordspacing

\bibitem{jeon2019analysis}
M.~Jeon, S.~Venkataraman, A.~Phanishayee, J.~Qian, W.~Xiao, and F.~Yang, ``Analysis of large-scale multi-tenant gpu clusters for dnn training workloads,'' in \emph{2019 USENIX Annual Technical Conference (USENIX ATC 19)}, 2019, pp. 947--960.

\bibitem{pytorch}
\BIBentryALTinterwordspacing
``Pytorch,'' 2024. [Online]. Available: \url{https://pytorch.org}
\BIBentrySTDinterwordspacing

\bibitem{deepspeed}
\BIBentryALTinterwordspacing
``Deepspeed,'' 2024. [Online]. Available: \url{https://www.deepspeed.ai/}
\BIBentrySTDinterwordspacing

\end{thebibliography}

\end{document}